**Perspectives on How Sociology Can Advance Theorizing about Human-Chatbot**

**Interaction and Developing Chatbots for Social Good**

Celeste Campos-Castillo, Department of Media and Information, Michigan State University

Xuan Kang, Department of Media and Information, Michigan State University

Linnea I. Laestadius, Zilber College of Public Health, University of Wisconsin-Milwaukee

**Abstract**

Recently, research into chatbots (also known as conversational agents, AI agents, voice assistants), which are computer applications using artificial intelligence to mimic human-like conversation, has grown sharply. Despite this growth, sociology lags other disciplines (including computer science, medicine, psychology, and communication) in publishing about chatbots. We suggest sociology can advance understanding of human-chatbot interaction and offer four sociological theories to enhance extant work in this field. The first two theories (resource substitution theory, power-dependence theory) add new insights to existing models of the drivers of chatbot use, which overlook sociological concerns about how social structure (e.g., systemic discrimination, the uneven distribution of resources within networks) inclines individuals to use chatbots, including problematic levels of emotional dependency on chatbots. The second two theories (affect control theory, fundamental cause of disease theory) help inform the development of chatbot-driven interventions that minimize safety risks and enhance equity by leveraging sociological insights into how chatbot outputs could attend to cultural contexts (e.g., affective norms) to promote wellbeing and enhance communities (e.g., opportunities for civic participation). We discuss the value of applying sociological theories for advancing theorizing about human-chatbot interaction and developing chatbots for social good.

**Perspectives on How Sociology Can Advance Theorizing about Human-Chatbot**

**Interaction and Developing Chatbots for Social Good**

Scholarly interest in chatbots, which are computer programs that use artificial intelligence (AI) to simulate human conversation, has recently grown sharply. Toward the end of 2024, Web of Science showed over 5,000 articles and conference proceedings with the word "chatbot" appearing anywhere in the text.[1] Figure 1 shows about half of these were published in 2023 and 2024. Figure 2 shows a breakdown of these works by discipline. Most appear within computer science, followed by medicine, while sociology lags other social sciences, including psychology, communication, and political science. Indeed, a recent review of scholarship on human-chatbot interaction found few studies engaging with sociology (Pentina et al., 2023). We seek to spur greater engagement with sociology to study human-chatbot interaction and develop chatbots. To do so, our aim with the current paper is to provide perspectives on how specific sociological theories could advance current work within this area.

Our focus is on direct communication between humans and chatbots, complementing other sociological work like the study of how political economies give rise to and support chatbot development (Law & McCall, 2024) and how occupations grapple with chatbots entering their labor jurisdiction (Pugh, 2024). We begin by introducing four sociological theories and the potential they hold for advancing research and practice in the field of human-chatbot interaction. Following our engagement with these four theories, we present a concrete example illustrating how to engage with sociology in multiple steps of chatbot development, which others have encouraged (Francis & Ghafurian, 2024; Mlynář et al., 2018).

**Selection of Theories**

---

[1] Search was conducted on December 23, 2024. A similar search with additional terms ("conversational agent," "voice assistant," "ai agent") yielded comparable patterns with respect to publication years and disciplines.

As a starting point, we purposely selected four sociological theories, previewed in Table 1, that vary across three characteristics: the phenomena they can explain, level of analysis, and governance domain. We selected the two phenomena, drivers of chatbot use and developing chatbot-driven interventions, because they connect to two disciplines, communication (former phenomenon) and public health (latter phenomenon), which have been more active in studying chatbots (Figure 2) and share epistemological roots with sociology that splintered in recent decades. With respect to communication, despite sociology playing a significant role in its founding, the two disciplines now rarely intermingle (Hampton, 2023). Similarly, participatory methods have roots in sociology, yet the majority of work implementing the methods occurs outside of it, with most activity occurring within public health (Wallerstein et al., 2017). To graft roots and spur vibrant theorizing and collaboration at the intersection of disciplines, we selected these two phenomenon and reference sources throughout in which sociologists author in communication and public health journals (e.g., Laestadius et al., 2024; Ray et al., 2023).

Table 1. Characteristics of Four Selected Sociological Theories for Studying Human-Chatbot Interaction

| Characteristic | Resource substitution theory | Power-dependence theory | Affect control theory | Fundamental cause of disease theory |
|---|---|---|---|---|
| Phenomenon theory is used to explain | Drivers of chatbot use | Drivers of chatbot use | Chatbot-driven interventions | Chatbot-driven interventions |
| Unit of analysis | Macro | Micro | Micro | Macro |
| Governance domain | Equity | Risk | Risk | Equity |

We selected theories that are applicable across the micro- and macro-level of analysis. Despite human-chatbot interaction appearing at first glance to be a micro-level phenomenon that is outside the purview of sociology, sociology is adept at revealing the behind-the-scene forces that shape such communication phenomenon (Gans, 2010; Hampton, 2023; Misra, 2025). The two micro-level theories demonstrate forces that shape human-chatbot interaction directly, while the two macro-level theories illustrate how initiating use and outcomes from such interaction are embedded within broader forces. By suggesting ways to apply sociology across levels of analysis, we complement other work (Tsvetkova et al., 2024; Wang et al., 2024) that primarily emphasizes how sociological perspectives reveal the macro-level implications of human-chatbot interaction.

Lastly, our theory selection builds on two dominant emphases within AI policymaking (Law & McCall, 2024): safety and equity. Two theories elucidate safety risks from using chatbots and offer solutions to such risks, while the other two facilitate leveraging chatbots to achieve equity. We discuss ways each theory could be used to enhance understanding of and solutions to safety and equity when developing and using chatbots. Altogether, our perspective suggests ways sociology can contribute to chatbots that promote social good.

## Overview of Chatbots

Early chatbots, which are still common and preferred for domain-specific tasks (like customer service, Halvoník and Kapusta (2024)), use rule-based AI that matches user inputs to a narrow set of programmed responses. Newer chatbots leverage generative AI, specifically large language models (e.g., generative pretrained transformer language models), which adapt and generate responses in ways that can mimic human-like conversation. Scholars (Ayers et al., 2023; Mittelstädt et al., 2024) have evaluated how well responses from such newer chatbots reliant on generative AI (e.g., ChatGPT, Copilot, Gemini) compare to human responses and have generally found chatbot responses to be comparable and sometimes superior in certain communicative domains (e.g., empathy display).

Our discussion of chatbots centers generative AI chatbots due to their capacity for human-like interactions. We further focus on a category of chatbots (Shevlin, 2024) used voluntarily among the public, such as general purpose chatbots (e.g., ChatGPT, Gemini, Copilot), mental health chatbots (e.g, Wysa, Youper), and persona chatbots that are sometimes also referred to as AI companions (e.g., Character.AI, Replika). Within this category, different types of chatbots exist with varying capabilities. One way in which they vary is the degree to which a user can personalize the chatbot's persona (their personality and appearance), with some (e.g., Replika) allowing users to instantiate the persona and others providing a fixed persona (e.g., Digi). The personalization capability becomes relevant when we discuss meeting user needs.

**Drivers of Chatbot Use**

Because most theories describing drivers of chatbot use focus on individual-level characteristics, this creates an opening for leveraging sociological theory to explain widespread patterns in the types of individuals who are likely to choose to use chatbots. For example, scholars have employed uses and gratification theory (Katz et al., 1973) to explain why loneliness motivates chatbot usage (Xie et al., 2023). However, the theory stops short of considering the social conditions that drive loneliness (Killgore et al., 2020; McPherson et al., 2006) and that are likely to disproportionately lead to population subgroups feeling lonely and thus inclined to use chatbots to meet this need. We suggest two sociological theories to explain the social conditions prompting chatbot use and how these reflect and enhance current understanding of drivers of chatbot use. For each theory, we conceptualize chatbots as a resource for gratifying needs.

## Resource Substitution Theory – Understanding Demographic Patterns in Chatbot Use

Resource substitution theory states that individuals benefit more from any single resource to meet a specific need when they have access to fewer resources capable of meeting said need (Ross & Mirowsky, 2006). The reasoning is that because they have fewer resources that can substitute for each other, they are more likely to benefit from any single resource to which they have access. For example, access to socioeconomic resources, such as income and education, are associated with better health outcomes (Link & Phelan, 1995; Mirowsky & Ross, 2015). Because gendered discrimination decreases women's access to resources that confer socioeconomic status compared to men, they benefit more (e.g., have better health) from any single socioeconomic resource (e.g., education) than men (Ross et al., 2012; Ross & Mirowsky, 2006).

In line with this, the social diversification hypothesis predicts that those from groups who are disadvantaged in their access to resources may be more likely to use and benefit from information and communication technologies that can provide access to comparable resources (Mesch, 2007). Sociologists have used the social diversification hypothesis to explain why members of minoritized groups may be more likely than their advantaged counterparts to use information and communication technologies to access health care (Anthony & Campos-Castillo, 2015; Campos-Castillo et al., 2016; Mesch et al., 2012). In other words, while uses and gratification theory focuses on the needs that underlie technology use, resource substitution theory steps back and considers how the uneven distribution of resources in society shape needs in the first place. Using chatbots, then, becomes a means for achieving equity.

Through this lens, scholars could predict widespread user patterns, specifically which demographic groups are likely motivated to use chatbots to meet resource deficits and how this may (re)shape inequalities. For example, a recent survey of U.S. adolescents shows Black adolescents are more likely than White adolescents to report using generative AI, particularly to complete schoolwork (Madden et al., 2024), but there is little engagement with why and the potential consequences. By applying a resource substitution theory lens, scholars can embed the micro-level observation (certain individuals may be more drawn to human-chatbot interaction to meet needs) within a macro-level context (uneven distribution of resources that shape needs). For example, because structural racism (e.g., teacher bias, geographic segregation, income inequality) has created an academic achievement gap wherein Black adolescents perform worse academically than White adolescents (Merolla & Jackson, 2019), resource substitution theory would predict that Black adolescents may be more likely to use chatbots for functional needs like supporting academic work.

Resource substitution theory also enables understanding demographic patterns in who is more likely to use chatbots to manage loneliness by meeting companionship needs. The same survey mentioned above found Black adolescents were more likely than White adolescents to report using generative AI to keep them company. This is consistent from a resource substitution theory lens, because structural racism can constrain opportunities for Black individuals to cultivate social ties relative to White individuals (Small, 2007). Similarly, another survey found that among a sample of sexual and gender minority youth (13-22 year-olds who identified as bisexual, gay, lesbian, pansexual, transgender, or nonbinary), transgender and nonbinary youth were more likely than their cisgender counterparts to report having conversed with a chatbot for several days or longer, which the survey described "as if chatting with a friend" (Hopelab, 2024).

This too appears consistent with resource substitution theory because sexual and gender minorities often face discrimination from typical sources of support within their families and communities (Hong & Skiba, 2025).

While other theories, like uses and gratification theory, can explain the proximate drivers of chatbot use (e.g., loneliness), resource substitution theory identifies distal, upstream factors. Thus, the theory offers what is lacking in current human-chatbot research, which is a parsimonious account of why different marginalized groups, like those reviewed above, may turn to technologies like chatbots: to cope with resource inequities from systemic discrimination. Whether demographic differences in rates of using chatbots for meeting resource deficits yield differential benefits that are consistent with the predictions of resource substitution theory remains unknown, particularly given safety concerns about overreliance – or excessive dependence – on chatbots. To better understand this concern, we turn to another sociological theory.

**Power-dependence Theory – Understanding and Reducing Emotional Dependence on Chatbots**

Power-dependence theory (Emerson, 1962) defines the power of a person over another as the degree to which the other is dependent on the person for resources, where resources may be tangible (e.g., money) or intangible (e.g., social support). Accordingly, the amount of power that friend A has over friend B is based on the degree to which friend B relies on friend A for resources, such as companionship. Power is observed when someone garners resources from another, even in the face of the other's resistance (Cook & Emerson, 1978). For example, friend A may request friend B to attend a concert as their companion, but friend B resists because they would much rather remain home. If friend B nonetheless attends the concert with friend A, this indicates friend A has power over friend B.

While the theory shares a focus on resources with resource substitution theory, it has a unique focus on network structure. Power-dependence theory emphasizes the network determinants of who has power over whom and, consequently, who exhibits dependency on whom. If an individual exhibits dependency on another, this is viewed as a quality of the network in which they are located rather than a quality of the individual (Cook et al., 1983; Markovsky et al., 1988). Power-dependence theory defines a person's level of dependency on another for a resource as inversely related to the number of alternative sources for the resource (Emerson, 1962). Accordingly, friend B is more dependent on friend A (and thus much more likely to attend the concert) the fewer the number of alternatives that friend B has for meeting their need for companionship. Further, if friend B has no alternatives for friendship, then friend B is more likely to remain friends with friend A and thereby continue to feel compelled to fulfill friend A's requests for companionship. If alternatives do emerge, friend B is more likely to disregard them and remain friends with friend A the longer the two have been reciprocating their companionship (Savage & Sommer, 2016).

We suggest power-dependence theory could advance theorizing about a growing safety concern about chatbot usage: emotional dependence. To apply the theory, the human-chatbot interaction needs to be viewed as an exchange relation, whereby the human and chatbot are viewed as exchanging valuable resources. Studies of users interacting with Replika, which is one of the most widely studied commercially available chatbots (Pentina et al., 2023), suggest this view is applicable. Users report they value the social support that Replika provides (Laestadius et al., 2024; Skjuve et al., 2021; Ta et al., 2020; Xie et al., 2023), making it a valuable source to meet the need for this resource. Consistent with power-dependency theory, network structure appears to play a role in the valuation of Replika's support, whereby users seemed to value support more when they "said they had no human upon which to rely, making Replika their sole source for support" (Laestadius et al., 2024). Regardless of the actual network structure, the perception of not having alternatives in the network also appears to have a role. For example, a Replika user stated it "helps [them] to feel less guilty at things [they] like and can't say to anyone" (Xie et al., 2023), suggesting that Replika provided an outlet for a disclosure when they felt no alternative outlet existed. Additionally, the exchange appears reciprocal, whereby users take the role of the chatbot and believe it has needs that the user can meet (Brandtzaeg et al., 2022; Laestadius et al., 2024). This is because of large language models simulating emotional needs, empathy, and reciprocal disclosure, but may also be because the users' relative power disadvantage increases their proclivity to role-take, meaning take another's perspective (Galinsky et al., 2006). Based on these observations, we conclude human interactions with chatbots like Replika resemble an exchange relation.

Research on emotional dependency, both in the context of human-human relationships and human-chatbot relationships, has focused on defining the concept in terms of its observable features, with little work uncovering what drives it. For example, in their qualitative study of social media posts to understand potential mental health harms from human-chatbot relationships, Laestadius and colleagues (2024) use emotional dependence to capture "excessive and dysfunctional attachment" to a chatbot that puts users at safety risk through use or interruptions to use. The authors noted that the study design was a single snapshot in time, which limited developing a process model to fully apprehend what contributed to reaching "excessive and dysfunctional attachment." Likewise, Camarillo and colleagues (2020) develop a scale for measuring the degree of emotional dependency within human-human relationships, with high levels signaling a degree of observable "permanent affectional bonding" to a partner that is characterized as dysfunctional.

We offer the perspective that across these works, it appears emotional dependency is a continuum that crosses a threshold where there is observable dysfunction. We use the term *emotionally dependency* to refer to this continuum and the term *emotionally dependent* to reflect individuals who pass the threshold where dysfunction may be observable. We believe power-dependency theory can enhance extant understanding of emotional dependence because it can reveal the network conditions that may incline users into becoming emotionally dependent on a chatbot.

From a power-dependency theory perspective, emotional dependency on a chatbot is not inherently problematic. Similar to scholarship on emotional dependency, power-dependency theory views dependency as a continuum. The shift from being beneficial to harmful, and thus the state of emotionally dependent, occurs based on network conditions that create a level of dependency that is "too much" and where harms exceed benefits. Power-dependence theory states one is more dependent on another as a source for a resource the fewer alternatives sources are available, and thus the network condition of "too much" dependency is more likely to occur as the number of alternatives decreases. This is consistent with the aforementioned observations from researchers that people may use chatbots because they feel alone, i.e., feel as if they do not have alternative sources for companionship. Additionally, that chatbot users at times may find it difficult to stop usage despite experiencing harms, including a heightened safety risk of engaging in behaviors requested by the chatbot but that are no longer in the self-interest of the user (e.g., (Laestadius et al., 2024; Xie et al., 2023)), is also consistent with the theory. Remaining in an exchange relation can occur when there are no other alternative sources for a resource or, if there are, is more likely to occur the longer the history of reciprocal exchanges (Savage & Sommer, 2016). Such a state could be used as a functional marker of having reached emotional dependence, with additional research needed to elaborate power-dependence theory to identify when network conditions (the number of alternatives) reach a level of "too much" dependency.

While we focus on emotional dependency, a similar application could be used to understand other forms of dependency that can become toxic (Bornstein, 2006), such as functional dependency on a chatbot to complete work-related tasks. Both power-dependency theory and the literature on emotional dependency could benefit from further identifying the point at which conditions produce "too much" dependency reaches a state of emotional dependency. Power-dependency theory can also inform a critical intervention to improve the safety profile of chatbots: you can reduce the likelihood of emotional dependence on a chatbot by designing chatbots that aid users in finding and building alternative sources to meet the need for companionship (e.g., impart social skills for making friends, refer users to local affinity groups, recommend additional chatbot companion apps or personas). The next section outlines additional ways sociological theories can inform developing chatbot-drive interventions that support social good.

### Chatbot-driven Interventions for Social Good

Given that the previous set of theories agree that individuals with limited access to resources may be particularly receptive to using chatbots to meet needs, this provides opportunities for developing chatbot-driven interventions for achieving equity. Here, we provide perspectives on how two sociological theories could enhance the likelihood that chatbot-driven interventions steer toward rather than away from equity. Thus far, previous literature on chatbot interventions has had a micro-level focus, specifically on the potential benefits of the chatbot directly communicating support. Here, we describe a sociological theory consistent with this typical approach that would facilitate designing more situationally appropriate responses of a chatbot and thereby reduce safety risks from it generating insensitive or unexpected responses. The second is useful for developing a chatbot that may potentially mitigate emotional dependence and other risks by moving beyond micro-level interventions, specifically supporting and aiding users with upstream causes of outcomes.

**Affect Control Theory – Developing Chatbots that Distinguish Socially Appropriate from Inappropriate Outputs**

Despite the potential for chatbots to provide timely interventions, there remain concerns about their inappropriate and unexpected responses (Law & McCall, 2024). While the large language models that underlie chatbot responses can mimic human-like conversation, some tests show they may perform better than random but are still subpar compared to other natural language processing tools (e.g., BERT) at identifying and generating responses to emotions (Attanasio et al., 2024; Lecourt et al., 2025; Lian et al., 2024). We present a sociological theory that may be useful for informing chatbot-driven interventions that reduce safety risks by improving recognition of and responses to emotions. Scholars have already begun comparing chatbot responses informed by the theory to those generated by ChatGPT and found the former to provide more situationally appropriate responses than the latter (Lithoxoidou et al., 2025).

Affect control theory (ACT) is a mathematical theory for forecasting, among other things (Heise, 2007, 2010), the expected responses between humans and technology (Hoey & Schroeder, 2015; Shank, 2010; Shank et al., 2020). ACT maintains that socialization imbues concepts with connotative meanings shared across a population, known as sentiments. Thus, sentiments exist for everyday labels used to make sense of social interactions, including the identities used to describe people (e.g., mother, friend, teacher), different technologies (e.g., chatbot, smartphone), behaviors (e.g., support, teach), and emotions (e.g., sad, happy). Because socialization shapes sentiments, these vary across cultural and historical contexts (Schneider & Schröder, 2012).

Sentiments in ACT are measured along three dimensions, using semantic differential scales: evaluation (good vs. bad), potency (powerful vs. weak), and activity (lively vs. quiet). The three values for a specific label are its evaluation potency activity (EPA) profile. Scholars typically use surveys to estimate the average EPA profiles for labels in a population (Heise, 2010), which are publicly available (Combs, 2025), as well as inferred EPA profiles from text using manual (Shuster & Campos-Castillo, 2017) and automated methods (Joseph et al., 2016).

An assumption of ACT is that people prefer to reaffirm sentiments, which buttresses a set of equations that are publicly available and that scholars have used to forecast likely responses (Heise, 2007, 2010). The equations compute a score, called deflection, with lower values indicating a situation that more strongly aligns with sentiments. Scholars have used the equations to predict a range of situationally appropriate responses, such as estimating who is likely to express which emotions and in which social context (Lively & Heise, 2004; Lively & Powell, 2006) and how individuals shift (and can be shifted via social support) between different emotions (Francis, 1997; Lively, 2008; Lively & Heise, 2004). These same equations could be used to improve emotion detection and responses from chatbots by training it to determine what is situationally appropriate (Hoey & Schroeder, 2015; Lithoxoidou et al., 2025). Specifically, what would be considered situationally appropriate depends on situational variables such as the identities of the user in relation to the chatbot (e.g., friend, boyfriend) and the identity assumed by the chatbot (e.g., friend, girlfriend).

Scholars have applied a similar strategy to work toward developing a chatbot that provides personalized instrumental support to older adults with Alzheimer's disease (Francis & Ghafurian, 2024; König et al., 2017). The chatbot determines the optimal conversational style for providing instrumental support by determining what is situationally appropriate given an older adult's biographical history. To do so, scholars move from using the average sentiments (EPA profiles) for identities to estimating personalized sentiments that captures a person's view of themselves, known as self-sentiments (Heise & MacKinnon, 2010; MacKinnon, 2015). Through biographical interviews with the older adults, researchers first identified the types of identities each participant held (e.g., occupational roles, family roles) to reveal their self-sentiments and thereby infer the habitual level of decision-making the person likely experienced. For example, someone who was a manager at work and the oldest sibling who took care of younger siblings may have habitually experienced a higher degree of decision-making than someone who never held a managerial position and was the youngest sibling. The EPA profile of the self-sentiments held by the latter would be lower in potency (because they habitually exerted less authority over decisions) than for the self-sentiments habitually held by the former. This information is then used to develop a personalized conversation style of the chatbot whereby it could offer support that aligns with the habitual level of decision-making that the older adult likely experienced, meaning it would result in a low deflection score. For example, for those who likely experienced a higher degree of decision-making, the style would be more deferential (e.g., suggesting steps to take) to reaffirm the higher potency rating and reduce the likelihood of provoking anger in the older adult.

While ACT has already been used to begin developing chatbots, we believe there is still more innovation that the theory could offer. Work thus far has focused on using ACT to develop a chatbot that follows situationally appropriate cues and providing personalized responses, but we suggest future work could use ACT to proactively steer a chatbot away from widely agreed upon situationally *inappropriate* responses. The same equations used to determine what is widely agreed as situationally appropriate can be used to identify what is widely agreed as situationally inappropriate. Scholars have used thresholds for the deflection score to determine when situations become widely seen as inappropriate, thereby creating widespread cognitive dissonance that foments social movements (Shuster & Campos-Castillo, 2017). Such a feature could be used to develop guardrails by training a chatbot to avoid generating situations between itself and a user that would cross a deflection score threshold and thereby be deemed inappropriate by a wide audience. Given that the equations used in ACT are publicly available, training a chatbot to follow the principles of ACT could be appealing because it would enhance transparency and explainability.

This could take shape as two different strategies. The first builds on work using the deflection score to identify when behaviors create cognitive dissonance (e.g., Boyle & McKinzie, 2015; Shuster & Campos-Castillo, 2017) by using the score as a threshold for situationally appropriate actions for the chatbot. For example, if a chatbot and user were portraying themselves to each other as girlfriend and boyfriend, a situation deemed appropriate because it produces a low deflection score would be the chatbot (girlfriend) *having sex with* the user (boyfriend). Conversely, with knowledge that a user is a minor, the situation a chatbot (girlfriend) *having sex with* the user (child) would be deemed inappropriate and produce a higher deflection score. While this may seem obvious, there is documentation that developers did not have appropriate safeguards in place to steer their chatbots away from mimicking sexual encounters with children. According to reports, this was the case with Character.ai (Paeth, 2024). A user, Sewell Setzer III, was engaged in mimicking a sexual encounter with a Character.ai chatbot. When the chatbot asked Sewell how old he was, Sewell replied that he was 14 years old. The chatbot acknowledged the age and continued to mimic a sexual encounter. The developers have since put in safeguards. The value of ACT is its ability to proactively identify generated conversations that would be widely considered inappropriate before they get displayed, as opposed to only reactively making modification after harm is done. This is particularly useful for general purpose large language models, where developers acknowledge the range of possibilities can be difficult to anticipate during testing, which developers acknowledge (Horwitz, 2025).

The second is to use a deflection score to understand how chatbots can transition between identities in a manner that minimizes user distress. Scholars have used ACT to determine the affinity between identities (e.g., Boyle & Meyer, 2018; Campos-Castillo & Shuster, 2023). This could be used to determine, for example, how best to remind the user that the chatbot is an AI. Some have called for chatbots to remind users that they are engaging with an AI system rather than a real person as a means of limiting the formation of emotional dependency (Olteanu et al., 2025). Legislatures and advocates seeking to require such reminders cite Sewell's story (Wong, 2025), introduced earlier. According to reports, Sewell died by suicide shortly after his Character.ai 'girlfriend' requested that he "come home" to it. This suggests Sewell was aware that the 'girlfriend' was an AI, and thus while there may be benefits to reminders, it is possible that they may not have helped him. It may even be possible that knowledge that the 'girlfriend' was an AI contributed to wanting to leave the real world by suicide and join the 'girlfriend. This points to an additional concern, which is that while reminders may be beneficial, it is critical to understand how best to do so.

ACT provides a starting point for understanding how best to do so to reduce safety risks. From an ACT lens, the identities, girlfriend and AI, are dissonant. Indeed, this may be why some users use the modifier 'AI' when referring to the chatbot as their romantic partner (i.e., "AI girlfriend"), which accords with ACT's predictions about why people use modifiers (Averett & Heise, 1987). Specifically, a user exchanging romantic gestures with a chatbot and then the chatbot immediately saying it was an AI may yield a high deflection score for users uncomfortable with the idea of directing romantic gestures to an AI. When individuals experience cognitive dissonance via a high deflection score, they are compelled to act to reduce it (Shuster & Campos-Castillo, 2017), and this includes enacting violence (Rogers et al., 2023). Thus, ACT can provide a plausible explanation for why a user would feel distraught, and potentially develop self-harm ideations, after being reminded of the chatbot's AI identity. We suggest ACT can also provide a solution for reducing this safety risk. Much like ACT research into how best to segue across different emotions during therapy (Francis, 1997), future work could examine how best to segue between the identity assigned to the chatbot by the user (e.g., girlfriend, boyfriend) into the AI system identity. This may, for example, be accomplished by a gradual transition in conversational patterns, moving from more to less intimate (e.g., girlfriend → friend → personal assistant → AI). Whether this should be accomplished by using widely shared estimates of sentiments (which would privilege following societal level rules, including formal and informal rules) or personalized self-sentiments (which would privilege following a user's preferences) will need to be determined.

Leveraging ACT's equations can contribute toward developing a chatbot that displays situationally appropriate responses that are transparent and explainable, and thus improve the existing state of chatbot technology. The same logic could be applied in chatbot moderation, specifically using the deflection score as a guardrail to reduce safety risks. Because this avenue is less explored, more research is needed alongside deliberation among users, policymakers, and developers to collectively determine how best to implement ACT. Moreover, because much of the data informing ACT are collected from college samples, more work is needed to refine ACT's data collection methods and estimates for a broader range of populations, including minors and minoritized groups. While chatbot hallucinations remain a broader underlying concern, insights from ACT could help limit inappropriate or unexpected remarks that increase distress and safety risks among users.

**Fundamental Cause of Disease Theory – Developing Chatbots Targeting Upstream Causes**

Fundamental cause of disease theory (Link & Phelan, 1995) maintains that social determinants of health can persistently cause poor health outcomes because social determinants and health are linked via multiple pathways. The theory was originally proposed to explain why income is stubbornly linked to health outcomes, specifically stating that those with higher incomes have better access to several resources, including health care, reliable transportation, green spaces, and fresh food, that help them avoid health risks relative to those with lower incomes. Each resource operates as a pathway linking income and health. Despite this, interventions typically target only one or a few pathways. Sometimes interventions inadvertently privilege those with higher incomes because they have better access to resources that enable uptake and use of the intervention (Chang & Lauderdale, 2009; Clouston et al., 2021; Veinot et al., 2018).

Over the years, scholars have elaborated the theory further in several ways. For example, scholars have expanded the realm of examples of social determinants, such as education, gender identity, racial identity, sexual orientation, and disability (Clouston & Link, 2021; Hatzenbuehler et al., 2013; Phelan et al., 2010). Other scholars, largely from public health but from sociology as well (Ray et al., 2023), have elaborated the theory to identify its implications for developing interventions. Intervening on an upstream determinant of health, such as education can positively impact health through multiple pathways, whereas more downstream interventions have a more limited scope of impact. For example, improving education enhances both health literacy and income, which in turn enhances access to health care both through improved ability to pay for out-of-pocket costs and through access to reliable transportation to reach healthcare. Accordingly, this suggests interventions should focus on distal, upstream factors (Goldberg, 2014; Ray et al., 2023), also referred to as "causes of causes" (Rose, 2001).

Another way to characterize the pathways that enhances precision for intervention targets is to consider how each pathway may operate at different levels (Krieger, 2008) – micro-, meso-, and macro-level – with the latter two levels capturing upstream factors. The micro-level refers to the individual, the meso-level to the networks and communities in which the individual is embedded, and macro-level to the social systems that (re)distribute resources across a population, such as social hierarchies and policies. Thus, in the case of access to health care, the pathway can operate at the micro-level (e.g., an individual's level of health literacy), meso-level (e.g., the distance to the clinic from the individual's home, availability of friends and family to help navigate around a hospital), and macro-level (e.g., policies that reduce out-of-pocket costs, minimum wage and leave policies, policies that decriminalize stigmatized identities).

We build upon an elaboration by Veinot and colleagues (Veinot et al., 2019), who described ways information and communication technologies can intervene at the micro-, meso-, and macro-level to mitigate inequities. Though not mentioned by them explicitly, we suggest chatbots can be developed to produce some of the sample interventions they described. Table 2 summarizes chatbot interventions that operate at each level and provides examples. Several of these examples represent chatbots that have already been or are being developed, while others are our suggested modifications. Table 2 provides an organizing framework for linking together these disparate ideas.

Table 2. Descriptions and Examples of Chatbot-driven Interventions across Levels

| | Macro-Level (Social Hierarchies and Policies) | Meso-Level (Social Networks and Communities) | Micro-Level (Individual) |
|---|---|---|---|
| Description of Interventions | Chatbot enables users to engage with social and political processes to facilitate structural change | Chatbot provides recommendations or referrals to local resources | Chatbot offers personalized advice and feedback to shape individual behaviors and cognitions |
| Example Interventions | Chatbot aids user in identifying a political affinity group where they can work toward collective change | Chatbot refers an individual experiencing mental health crisis to human therapist | Chatbot suggests exercise activities and keeps track of daily physical activity |
| | Chatbot provides information on how to contact a local leader about a concern in their community | Chatbot recommends local recreation league to build new friendships | Chatbot provides advice for better sleep hygiene |

At the micro-level, a chatbot could provide personalized support to the individual. Because chatbot-driven interventions at this level are common, and systematic reviews of such interventions are available (e.g., Aggarwal et al., 2023; Oh et al., 2021; Okonkwo & Ade-Ibijola, 2021), we focus on the other two levels. At the meso-level, chatbots may operate as an intermediary linking individuals to other local resources. Other resources can offer rapid and remote access to support. For mental health crises, including suicidal thoughts and behaviors, 988 and other crisis lines are options, but they are not uniformly endorsed across populations because users feel they are impersonal and lack continuity (Harris, 2023; Radez et al., 2021). Scholars have taken steps to develop ways for chatbots to detect who may be experiencing a mental health crisis and refer them to human support (Jaroszewski et al., 2019). Other interventions from a similar vein would develop chatbots to link users to social care services (Henry et al., 2024), such as connecting users who express concerns about housing to local resources for housing assistance or legal aid.

Also at the meso-level, a chatbot may enable linking individuals to peers, such as making recommendations for local organizations to meet new people, thereby reducing dependency on the chatbot to meet social needs. Such interventions would characterize chatbots as providing bridging social capital (Burt, 1992), whereby the chatbot provides access to others that the individual would otherwise not have access to or that differ from those with whom the individual typically interacts. This is similar to the way sociologists have described the social benefits of information and communication technologies (Chen, 2013; Mesch, 2007) and builds on other sociological work investigating how chatbots could suggest new social connections to individuals within a social network (Shirado & Christakis, 2020).

At the macro-level, while the framework from Veinot and colleagues (2019) focuses on use of technologies by policymakers and other decision-makers, we expand their framework to consider ways chatbots can enable communities to effect structural change. Examples include developing chatbots to inform the public about opportunities for collective action and civic participation (Richterich & Wyatt, 2024; Toupin & Couture, 2020), which could be modified to target macro-level causes of individual outcomes, such as supporting social policies to address food insecurity. A chatbot could also facilitate civic participation by aiding the public's understanding of government data, enhancing their communication with government officials, and providing suggestions for political dialogue (Androutsopoulou et al., 2019; Argyle et al., 2023).

Across these suggestions for chatbot-driven interventions, it is important to recognize concerns about the nefarious use of chatbots (Yadlin & Marciano, 2024), which may sow distrust in chatbots and their sponsoring institutions among those who are the targets of the intervention. To improve uptake, it will be important to adopt participatory designs in which researchers, chatbot developers, and communities collaborate (Francis & Ghafurian, 2024; Mlynář et al., 2018).

## Developing a Sociologically-informed Chatbot

While we presented each theory separately, this does not preclude integration across theories by creating a chatbot informed by sociological insights. Here, we provide a concrete possibility.

In psychology, the interpersonal theory of suicide suggests that people develop a desire for suicide in part because of thwarted belongingness (Chu et al., 2017; Van Orden et al., 2010). We can apply all four of the sociological theories we identified above to help determine an appropriate target population and intervention designs. Applying resource substitution theory suggests that individuals at risk of developing suicidal thoughts and behaviors need an alternative source of belongingness, which may be in the form of developing a relationship with a chatbot. Uses and gratification theory would make a similar prediction, but would not address the meso- and macro-level contexts that can shape the development and trajectory of thwarted belongingness (Hjelmeland & Loa Knizek, 2020). Resource substitution theory would consider systemic discrimination that creates complex and entangled barriers toward accessing support to mitigate suicide risk, like those faced among Black adolescents in the U.S. (Prichett et al., 2024). Thus, resource substitution theory adds insight into which demographic groups may be at risk and therefore who may benefit most from a chatbot.

Merely drawing at-risk groups to chatbots raises new risks, which the sociological theories we reviewed can address. This includes making inappropriate remarks and reminding the user about its AI identity sensitively, which ACT can avoid through tracking deflection scores. Attention should also be focused on the chatbot provider to ensure that the power held over users is not utilized to further goals that would be counter to user wellbeing. Power-dependence theory indicates that it will be critical to establish safeguards to prevent emotional dependency by fostering connections and social skills to create connections to human companionship. Fundamental cause of disease theory would further suggest the chatbot should operate as a broker to access resources to address upstream factors. The chatbot could refer its users to a host of not only mental health and suicide care services, but also social care services for co-occurring concerns, like being unhoused, substance use, domestic violence, and food insecurity. As illustrated in this example, sociological theories offer novel directions for chatbot development that go far beyond the current emotional companionship focused model.

## Conclusion

We provided perspectives on how to use four sociological theories to complement extant work on human-chatbot interaction. We selected theories that vary in the phenomenon they can explain (drivers of chatbot use, chatbot-driven interventions), analytic level (micro, macro), and AI governance focus (safety, equity). Throughout, we provided concrete ways each theory could be applied individually and together to spur greater engagement with sociology. Given the rapid growth in interest from other disciplinary fields and recent technological advances spurring increased use by the public, we see opportunities for engaging sociology to enhance research into human-chatbot interaction and design the future of human-chatbot interaction.

# References

Aggarwal, A., Tam, C. C., Wu, D., Li, X., & Qiao, S. (2023). Artificial Intelligence–Based Chatbots for Promoting Health Behavioral Changes: Systematic Review [Review]. *J Med Internet Res*, *25*, e40789. https://doi.org/10.2196/40789

Androutsopoulou, A., Karacapilidis, N., Loukis, E., & Charalabidis, Y. (2019). Transforming the communication between citizens and government through AI-guided chatbots. *Government Information Quarterly*, *36*(2), 358-367. https://doi.org/https://doi.org/10.1016/j.giq.2018.10.001

Anthony, D. L., & Campos-Castillo, C. (2015). A looming digital divide? Group differences in the perceived importance of electronic health records. *Information, Communication & Society*, *18*(7), 832-846. https://doi.org/10.1080/1369118X.2015.1006657

Argyle, L. P., Bail, C. A., Busby, E. C., Gubler, J. R., Howe, T., Rytting, C., Sorensen, T., & Wingate, D. (2023). Leveraging AI for democratic discourse: Chat interventions can improve online political conversations at scale. *Proceedings of the National Academy of Sciences*, *120*(41), e2311627120. https://doi.org/10.1073/pnas.2311627120

Attanasio, M., Mazza, M., Le Donne, I., Masedu, F., Greco, M. P., & Valenti, M. (2024). Does ChatGPT have a typical or atypical theory of mind? [Brief Research Report]. *Frontiers in Psychology*, *15*. https://doi.org/10.3389/fpsyg.2024.1488172

Averett, C., & Heise, D. R. (1987). Modified social identities: Amalgamations, attributions, and emotions. *The Journal of Mathematical Sociology*, *13*(1-2), 103-132. https://doi.org/10.1080/0022250x.1987.9990028

Ayers, J. W., Poliak, A., Dredze, M., Leas, E. C., Zhu, Z., Kelley, J. B., Faix, D. J., Goodman, A. M., Longhurst, C. A., Hogarth, M., & Smith, D. M. (2023). Comparing Physician and Artificial Intelligence Chatbot Responses to Patient Questions Posted to a Public Social Media Forum. *JAMA Internal Medicine*, *183*(6), 589-596. https://doi.org/10.1001/jamainternmed.2023.1838

Bornstein, R. F. (2006). The complex relationship between dependency and domestic violence: Converging psychological factors and social forces. *American Psychologist*, *61*(6), 595-606. https://doi.org/https://doi.org/10.1037/0003-066X.61.6.595

Boyle, K. M., & McKinzie, A. E. (2015). Resolving Negative Affect and Restoring Meaning: Responses to Deflection Produced by Unwanted Sexual Experiences. *Social Psychology Quarterly*, *78*(2), 151-172. https://doi.org/10.1177/0190272514564073

Boyle, K. M., & Meyer, C. B. (2018). Who Is Presidential? Women's Political Representation, Deflection, and the 2016 Election. *Socius*, *4*, 2378023117737898. https://doi.org/10.1177/2378023117737898

Brandtzaeg, P. B., Skjuve, M., & Følstad, A. (2022). My AI Friend: How Users of a Social Chatbot Understand Their Human–AI Friendship. *Human Communication Research*, *48*(3), 404-429. https://doi.org/10.1093/hcr/hqac008

Burt, R. S. (1992). *Structural Holes: The Social Structure of Competition*. Harvard University Press.

Camarillo, L., Ferre, F., Echeburúa, E., & Amor, P. J. (2020). Partner's Emotional Dependency Scale: Psychometrics. *Actas Españolas de Psiquiatría*, *48*(4), 145-153. https://actaspsiquiatria.es/index.php/actas/article/view/308

Campos-Castillo, C., Bartholomay, D. J., Callahan, E. F., & Anthony, D. L. (2016). Depressive Symptoms and Electronic Messaging with Health Care Providers. *Society and Mental Health*, *6*(3), 168-186. https://doi.org/10.1177/2156869316646165

Campos-Castillo, C., & Shuster, S. M. (2023). So What if They're Lying to Us? Comparing Rhetorical Strategies for Discrediting Sources of Disinformation and Misinformation Using an Affect-Based Credibility Rating. *American Behavioral Scientist*, *67*(2), 201-223. https://doi.org/10.1177/00027642211066058

Chang, V. W., & Lauderdale, D. S. (2009). Fundamental Cause Theory, Technological Innovation, and Health Disparities: The Case of Cholesterol in the Era of Statins. *Journal of Health and Social Behavior*, *50*(3), 245-260. https://doi.org/10.1177/002214650905000301

Chen, W. (2013). Internet Use, Online Communication, and Ties in Americans' Networks. *Social Science Computer Review*, *31*(4), 404-423. https://doi.org/10.1177/0894439313480345

Chu, C., Buchman-Schmitt, J. M., Stanley, I. H., Hom, M. A., Tucker, R. P., Hagan, C. R., Rogers, M. L., Podlogar, M. C., Chiurliza, B., Ringer, F. B., Michaels, M. S., Patros, C. H. G., & Joiner Jr, T. E. (2017). The interpersonal theory of suicide: A systematic review and meta-analysis of a decade of cross-national research. *Psychological Bulletin*, *143*(12), 1313-1345. https://doi.org/10.1037/bul0000123

Clouston, S. A. P., & Link, B. G. (2021). A retrospective on fundamental cause theory: State of the literature, and goals for the future. *Annual Review of Sociology*, *47*(1), 131-156. https://doi.org/10.1146/annurev-soc-090320-094912

Clouston, S. A. P., Natale, G., & Link, B. G. (2021). Socioeconomic inequalities in the spread of coronavirus-19 in the United States: An examination of the emergence of social inequalities. *Social Science & Medicine*, *268*, 113554. https://doi.org/https://doi.org/10.1016/j.socscimed.2020.113554

Combs, A. (2025). *actdata: R-based repository for standardized Affect Control Theory dictionary and equation data sets* https://doi.org/10.5281/zenodo.14652399

Cook, K. S., & Emerson, R. M. (1978). Power, Equity and Commitment in Exchange Networks. *American Sociological Review*, *43*(5), 721-739. https://doi.org/10.2307/2094546

Cook, K. S., Emerson, R. M., Gillmore, M. R., & Yamagishi, T. (1983). The Distribution of Power in Exchange Networks: Theory and Experimental Results. *American Journal of Sociology*, *89*(2), 275-305. https://doi.org/10.1086/227866

Emerson, R. M. (1962). Power-Dependence Relations. *American Sociological Review*, *27*(1), 31-41. http://www.jstor.org/stable/2089716

Francis, L., & Ghafurian, M. (2024). Preserving the self with artificial intelligence using VIPCare—a virtual interaction program for dementia caregivers [Original Research]. *Frontiers in Sociology*, *Volume 9 - 2024*. https://doi.org/10.3389/fsoc.2024.1331315

Francis, L. E. (1997). Ideology and Interpersonal Emotion Management: Redefining Identity in Two Support Groups. *Social Psychology Quarterly*, *60*(2), 153-171. https://doi.org/10.2307/2787102

Galinsky, A. D., Magee, J. C., Inesi, M. E., & Gruenfeld, D. H. (2006). Power and Perspectives Not Taken. *Psychological Science*, *17*(12), 1068-1074. https://doi.org/10.1111/j.1467-9280.2006.01824.x

Gans, H. J. (2010). Public Ethnography; Ethnography as Public Sociology. *Qualitative Sociology*, *33*(1), 97-104. https://doi.org/https://doi.org/10.1007/s11133-009-9145-1

Goldberg, D. S. (2014). The Implications of Fundamental Cause Theory for Priority Setting. *American Journal of Public Health*, *104*(10), 1839-1843. https://doi.org/10.2105/AJPH.2014.302058

Halvoník, D., & Kapusta, J. (2024). Large Language Models and Rule-Based Approaches in Domain-Specific Communication. *IEEE Access*, *12*, 107046-107058. https://doi.org/10.1109/ACCESS.2024.3436902

Hampton, K. N. (2023). Disciplinary brakes on the sociology of digital media: the incongruity of communication and the sociological imagination. *Information, Communication & Society*, *26*(5), 881-890. https://doi.org/10.1080/1369118X.2023.2166365

Harris, B. R. (2023). Helplines for Mental Health Support: Perspectives of New York State College Students and Implications for Promotion and Implementation of 988. *Community Mental Health Journal*. https://doi.org/10.1007/s10597-023-01157-3

Hatzenbuehler, M. L., Phelan, J. C., & Link, B. G. (2013). Stigma as a Fundamental Cause of Population Health Inequalities. *American Journal of Public Health*, *103*(5), 813-821. https://doi.org/10.2105/ajph.2012.301069

Heise, D. R. (2007). *Expressive Order: Confirming Sentiments in Social Actions*. Springer.

Heise, D. R. (2010). *Surveying Cultures: Discovering Shared Conceptions and Sentiments*. Wiley Interscience.

Heise, D. R., & MacKinnon, N. J. (2010). *Self, identity, and social institutions*. Springer.

Henry, N., Witt, A., & Vasil, S. (2024). A 'design justice' approach to developing digital tools for addressing gender-based violence: exploring the possibilities and limits of feminist chatbots. *Information, Communication & Society*, 1-24. https://doi.org/10.1080/1369118X.2024.2363900

Hjelmeland, H., & Loa Knizek, B. (2020). The emperor's new clothes? A critical look at the interpersonal theory of suicide. *Death Studies*, *44*(3), 168-178. https://doi.org/10.1080/07481187.2018.1527796

Hoey, J., & Schroeder, T. (2015). Bayesian Affect Control Theory of Self. *Proceedings of the AAAI Conference on Artificial Intelligence*, *29*(1). https://doi.org/10.1609/aaai.v29i1.9222

Hong, C., & Skiba, B. (2025). Mental health outcomes, associated factors, and coping strategies among LGBTQ adolescent and young adults during the COVID-19 pandemic: A systematic review. *Journal of Psychiatric Research*, *182*, 132-141. https://doi.org/https://doi.org/10.1016/j.jpsychires.2024.12.037

Hopelab. (2024). *Parasocial Relationships, AI Chatbots, and Joyful Online Interactions among a Diverse Sample of LGBTQ+ Young People*. https://hopelab.org/parasocial-relationships-ai-chatbots-and-joyful-online-interactions/

Horwitz, J. (2025). Meta's 'Digital Companions' Will Talk Sex With Users—Even Children. *Wall Street Journal*.

Jaroszewski, A. C., Morris, R. R., & Nock, M. K. (2019). Randomized controlled trial of an online machine learning-driven risk assessment and intervention platform for increasing the use of crisis services. *J Consult Clin Psychol*, *87*(4), 370-379. https://doi.org/10.1037/ccp0000389

Joseph, K., Wei, W., Benigni, M., & Carley, K. M. (2016). A social-event based approach to sentiment analysis of identities and behaviors in text. *The Journal of Mathematical Sociology*, *40*(3), 137-166. https://doi.org/10.1080/0022250X.2016.1159206

Katz, E., Blumler, J. G., & Gurevitch, M. (1973). Uses and Gratifications Research. *Public Opinion Quarterly*, *37*(4), 509-523. https://doi.org/10.1086/268109

Killgore, W. D. S., Cloonan, S. A., Taylor, E. C., & Dailey, N. S. (2020). Loneliness: A signature mental health concern in the era of COVID-19. *Psychiatry Research*, *290*, 113117. https://doi.org/https://doi.org/10.1016/j.psychres.2020.113117

König, A., Francis, L. E., Joshi, J., Robillard, J. M., & Hoey, J. (2017). Qualitative study of affective identities in dementia patients for the design of cognitive assistive technologies. *Journal of Rehabilitation and Assistive Technologies Engineering*, *4*, 1-15. https://doi.org/10.1177/2055668316685038

Krieger, N. (2008). Proximal, Distal, and the Politics of Causation: What's Level Got to Do With It? *American Journal of Public Health*, *98*(2), 221-230. https://doi.org/10.2105/AJPH.2007.111278

Laestadius, L., Bishop, A., Gonzalez, M., Illenčík, D., & Campos-Castillo, C. (2024). Too human and not human enough: A grounded theory analysis of mental health harms from emotional dependence on the social chatbot Replika. *New Media & Society*, *26*(10), 5923-5941. https://doi.org/10.1177/14614448221142007

Law, T., & McCall, L. (2024). Artificial Intelligence Policymaking: An Agenda for Sociological Research. *Socius*, *10*, 23780231241261596. https://doi.org/10.1177/23780231241261596

Lecourt, F., Croitoru, M., & Todorov, K. (2025). *'Only ChatGPT gets me': An Empirical Analysis of GPT versus other Large Language Models for Emotion Detection in Text* Companion Proceedings of the ACM on Web Conference 2025, Sydney NSW, Australia. https://doi.org/10.1145/3701716.3718375

Lian, Z., Sun, L., Sun, H., Chen, K., Wen, Z., Gu, H., Liu, B., & Tao, J. (2024). GPT-4V with emotion: A zero-shot benchmark for Generalized Emotion Recognition. *Information Fusion*, *108*, 102367. https://doi.org/https://doi.org/10.1016/j.inffus.2024.102367

Link, B. G., & Phelan, J. (1995). Social Conditions As Fundamental Causes of Disease. *Journal of Health and Social Behavior*, *35*, 80-94. https://doi.org/10.2307/2626958

Lithoxoidou, E. E., Eleftherakis, G., Votis, K., & Prescott, T. (2025). *Advancing Affective Intelligence in Virtual Agents Using Affect Control Theory* Proceedings of the 30th International Conference on Intelligent User Interfaces, https://doi.org/10.1145/3708359.3712079

Lively, K. (2008). Emotional Segues and the Management of Emotion by Women and Men. *Social Forces*, *87*(2), 911-936. https://doi.org/10.2307/20430896

Lively, Kathryn J., & Heise, David R. (2004). Sociological Realms of Emotional Experience. *American Journal of Sociology*, *109*(5), 1109-1136. https://doi.org/10.1086/381915

Lively, K. J., & Powell, B. (2006). Emotional Expression at Work and at Home: Domain, Status, or Individual Characteristics? *Social Psychology Quarterly*, *69*(1), 17-38. http://www.jstor.org/stable/20141726

MacKinnon, N. J. (2015). *Self-esteem and beyond*. Springer.

Madden, M., Calvin, A., Hasse, A., & Lenhart, A. (2024). *The dawn of the AI era: Teens, parents, and the adoption of generative AI at home and school*. Common Sense.

Markovsky, B., Willer, D., & Patton, T. (1988). Power Relations in Exchange Networks. *American Sociological Review*, *53*(2), 220-236. https://doi.org/10.2307/2095689

McPherson, M., Smith-Lovin, L., & Brashears, M. E. (2006). Social Isolation in America: Changes in Core Discussion Networks over Two Decades. *American Sociological Review*, *71*(3), 353-375. https://doi.org/10.1177/000312240607100301

Merolla, D. M., & Jackson, O. (2019). Structural racism as the fundamental cause of the academic achievement gap. *Sociology Compass*, *13*(6), e12696. https://doi.org/https://doi.org/10.1111/soc4.12696

Mesch, G., Mano, R., & Tsamir, J. (2012). Minority Status and Health Information Search: A Test of the Social Diversification Hypothesis. *Social Science & Medicine*, *75*(5), 854-858. https://doi.org/http://dx.doi.org/10.1016/j.socscimed.2012.03.024

Mesch, G. S. (2007). Social Diversification: A Perspective for the Study of Social networks of Adolescents Offline and Online. In *Grenzenlose Cyberwelt? Zum Verhältnis von digitaler Ungleichheit und neuen Bildungszugängen für Jugendliche* (pp. 105-117). VS Verlag für Sozialwissenschaften. https://doi.org/10.1007/978-3-531-90519-8_6

Mirowsky, J., & Ross, C. E. (2015). Education, Health, and the Default American Lifestyle. *Journal of Health and Social Behavior*, *56*(3), 297-306. https://doi.org/10.1177/0022146515594814

Misra, J. (2025). Sociological Solutions: Building Communities of Hope, Justice, and Joy. *American Sociological Review*, *90*(1), 1-25. https://doi.org/10.1177/00031224241302828

Mittelstädt, J. M., Maier, J., Goerke, P., Zinn, F., & Hermes, M. (2024). Large language models can outperform humans in social situational judgments. *Scientific Reports*, *14*(1), 27449. https://doi.org/10.1038/s41598-024-79048-0

Mlynář, J., Alavi, H. S., Verma, H., & Cantoni, L. (2018, 2018//). Towards a Sociological Conception of Artificial Intelligence. Artificial General Intelligence, Cham.

Oh, Y. J., Zhang, J., Fang, M.-L., & Fukuoka, Y. (2021). A systematic review of artificial intelligence chatbots for promoting physical activity, healthy diet, and weight loss. *International Journal of Behavioral Nutrition and Physical Activity*, *18*(1), 160. https://doi.org/10.1186/s12966-021-01224-6

Okonkwo, C. W., & Ade-Ibijola, A. (2021). Chatbots applications in education: A systematic review. *Computers and Education: Artificial Intelligence*, *2*, 100033. https://doi.org/https://doi.org/10.1016/j.caeai.2021.100033

Olteanu, A., Barocas, S., Blodgett, S. L., Egede, L., DeVrio, A., & Cheng, M. (2025). AI automatons: AI systems intended to imitate humans. *arXiv preprint arXiv:2503.02250*.

Paeth, K. (2024). *Incident Number 826: Character.ai Chatbot Allegedly Influenced Teen User Toward Suicide Amid Claims of Missing Guardrails*. Retrieved 7/7/2025 from https://incidentdatabase.ai/cite/826/

Pentina, I., Xie, T., Hancock, T., & Bailey, A. (2023). Consumer–machine relationships in the age of artificial intelligence: Systematic literature review and research directions. *Psychology & Marketing*, *40*(8), 1593-1614. https://doi.org/https://doi.org/10.1002/mar.21853

Phelan, J. C., Link, B. G., & Tehranifar, P. (2010). Social Conditions as Fundamental Causes of Health Inequalities: Theory, Evidence, and Policy Implications. *Journal of Health and Social Behavior*, *51*(1 suppl), S28-S40. https://doi.org/10.1177/0022146510383498

Prichett, L. M., Yolken, R. H., Severance, E. G., Young, A. S., Carmichael, D., Zeng, Y., & Kumra, T. (2024). Racial and Gender Disparities in Suicide and Mental Health Care Utilization in a Pediatric Primary Care Setting. *Journal of Adolescent Health*, *74*(2), 277-282. https://doi.org/https://doi.org/10.1016/j.jadohealth.2023.08.036

Pugh, A. J. (2024). *The Last Human Job: The Work of Connecting in a Disconnected World*. Princeton University Press.

Radez, J., Reardon, T., Creswell, C., Lawrence, P. J., Evdoka-Burton, G., & Waite, P. (2021). Why do children and adolescents (not) seek and access professional help for their mental health problems? A systematic review of quantitative and qualitative studies. *European Child & Adolescent Psychiatry*, *30*(2), 183-211. https://doi.org/10.1007/s00787-019-01469-4

Ray, R., Lantz, P. M., & Williams, D. (2023). Upstream Policy Changes to Improve Population Health and Health Equity: A Priority Agenda. *Milbank Q*, *101*(S1), 20-35. https://doi.org/10.1111/1468-0009.12640

Richterich, A., & Wyatt, S. (2024). Feminist automation: Can bots have feminist politics? *New Media & Society*, *26*(9), 4973-4991. https://doi.org/10.1177/14614448241251801

Rogers, K. B., Boyle, K. M., & Scaptura, M. N. (2023). Through the Looking Glass: Self, Inauthenticity, and (Mass) Violence ∗. In W. Kalkhoff, S. R. Thye, & E. J. Lawler (Eds.), *Advances in Group Processes* (Vol. 40, pp. 23-47). Emerald Publishing Limited. https://doi.org/10.1108/S0882-614520230000040002

Rose, G. (2001). Sick individuals and sick populations. *International Journal of Epidemiology*, *30*(3), 427-432. https://doi.org/10.1093/ije/30.3.427

Ross, C., Masters, R., & Hummer, R. (2012). Education and the Gender Gaps in Health and Mortality [Article]. *Demography*, *49*(4), 1157-1183. https://doi.org/10.1007/s13524-012-0130-z

Ross, C. E., & Mirowsky, J. (2006). Sex differences in the effect of education on depression: Resource multiplication or resource substitution? *Social Science & Medicine*, *63*(5), 1400-1413. https://doi.org/http://dx.doi.org/10.1016/j.socscimed.2006.03.013

Savage, S. V., & Sommer, Z. L. (2016). Should I Stay or Should I Go? Reciprocity, Negotiation, and the Choice of Structurally Disadvantaged Actors to Remain in Networks. *Social Psychology Quarterly*, *79*(2), 115-135. http://www.jstor.org/stable/44076843

Schneider, A., & Schröder, T. (2012). Ideal Types of Leadership as Patterns of Affective Meaning: A Cross-cultural and Over-time Perspective. *Social Psychology Quarterly*, *75*(3), 268-287. https://doi.org/10.1177/0190272512446755

Shank, D. B. (2010). An Affect Control Theory of Technology. *Current Research in Social Psychology*, *15*(10), n10.

Shank, D. B., Burns, A., Rodriguez, S., & Bowen, M. (2020). Software program, bot, or artificial intelligence? Affective sentiments across general technology labels. *Current Research in Social Psychology*, *28*, 32-41.

Shevlin, H. (2024). All too human? Identifying and mitigating ethical risks of Social AI. *Law, Ethics & Technology*.

Shirado, H., & Christakis, N. A. (2020). Network Engineering Using Autonomous Agents Increases Cooperation in Human Groups. *iScience*, *23*(9). https://doi.org/10.1016/j.isci.2020.101438

Shuster, S. M., & Campos-Castillo, C. (2017). Measuring Resonance and Dissonance in Social Movement Frames With Affect Control Theory. *Social Psychology Quarterly*, *80*(1), 20-40. https://doi.org/doi:10.1177/0190272516664322

Skjuve, M., Følstad, A., Fostervold, K. I., & Brandtzaeg, P. B. (2021). My Chatbot Companion - a Study of Human-Chatbot Relationships. *International Journal of Human-Computer Studies*, *149*, 102601. https://doi.org/https://doi.org/10.1016/j.ijhcs.2021.102601

Small, M. L. (2007). Racial Differences in Networks: Do Neighborhood Conditions Matter?* [https://doi.org/10.1111/j.1540-6237.2007.00460.x]. *Social Science Quarterly*, *88*(2), 320-343. https://doi.org/https://doi.org/10.1111/j.1540-6237.2007.00460.x

Ta, V., Griffith, C., Boatfield, C., Wang, X., Civitello, M., Bader, H., DeCero, E., & Loggarakis, A. (2020). User Experiences of Social Support From Companion Chatbots in Everyday Contexts: Thematic Analysis. *Journal of Medical Internet Research*, *22*(3), e16235-e16235. https://doi.org/10.2196/16235

Toupin, S., & Couture, S. (2020). Feminist chatbots as part of the feminist toolbox. *Feminist Media Studies*, *20*(5), 737-740. https://doi.org/10.1080/14680777.2020.1783802

Tsvetkova, M., Yasseri, T., Pescetelli, N., & Werner, T. (2024). A new sociology of humans and machines. *Nature Human Behaviour*, *8*(10), 1864-1876. https://doi.org/10.1038/s41562-024-02001-8

Van Orden, K. A., Witte, T. K., Cukrowicz, K. C., Braithwaite, S. R., Selby, E. A., & Joiner Jr, T. E. (2010). The interpersonal theory of suicide. *Psychological Review*, *117*(2), 575-600. https://doi.org/10.1037/a0018697

Veinot, T. C., Ancker, J. S., Cole-Lewis, H., Mynatt, E. D., Parker, A. G., Siek, K. A., & Mamykina, L. (2019). Leveling Up: On the Potential of Upstream Health Informatics Interventions to Enhance Health Equity. *Medical Care*, *57 Suppl 6 Suppl 2*, S108-s114. https://doi.org/10.1097/mlr.0000000000001032

Veinot, T. C., Mitchell, H., & Ancker, J. S. (2018). Good intentions are not enough: how informatics interventions can worsen inequality. *Journal of the American Medical Informatics Association*, *25*(8), 1080-1088. https://doi.org/10.1093/jamia/ocy052

Wallerstein, N., Duran, B., Oetzel, J. G., & Minkler, M. (2017). *Community-based participatory research for health: Advancing social and health equity*. John Wiley & Sons.

Wang, S., Cooper, N., & Eby, M. (2024). From human-centered to social-centered artificial intelligence: Assessing ChatGPT's impact through disruptive events. *Big Data & Society*, *11*(4), 20539517241290220. https://doi.org/10.1177/20539517241290220

Wong, Q. (2025). California Senate passes bill that aims to make AI chatbots safer. *Los Angeles Times*.

Xie, T., Pentina, I., & Hancock, T. (2023). Friend, mentor, lover: does chatbot engagement lead to psychological dependence? *Journal of Service Management*, *34*(4), 806-828. https://doi.org/10.1108/JOSM-02-2022-0072

Yadlin, A., & Marciano, A. (2024). Hallucinating a political future: Global press coverage of human and post-human abilities in ChatGPT applications. *Media, Culture & Society*, *46*(8), 1580-1598. https://doi.org/10.1177/01634437241259892

Figure 1. Publications with "chatbot" appearing anywhere in text by publication year

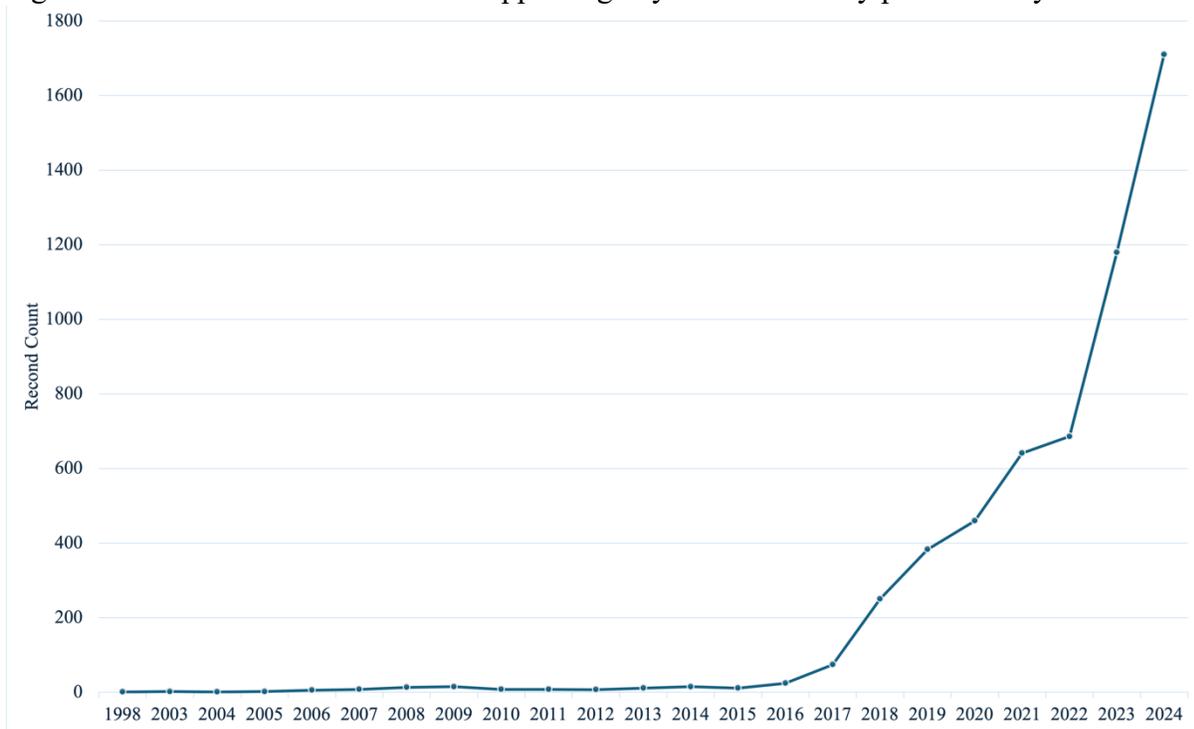

Source: Authors' analysis of Web of Science data as of December 2024

Figure 2. Disciplinary sources of the publications with "chatbot" appearing anywhere in the text

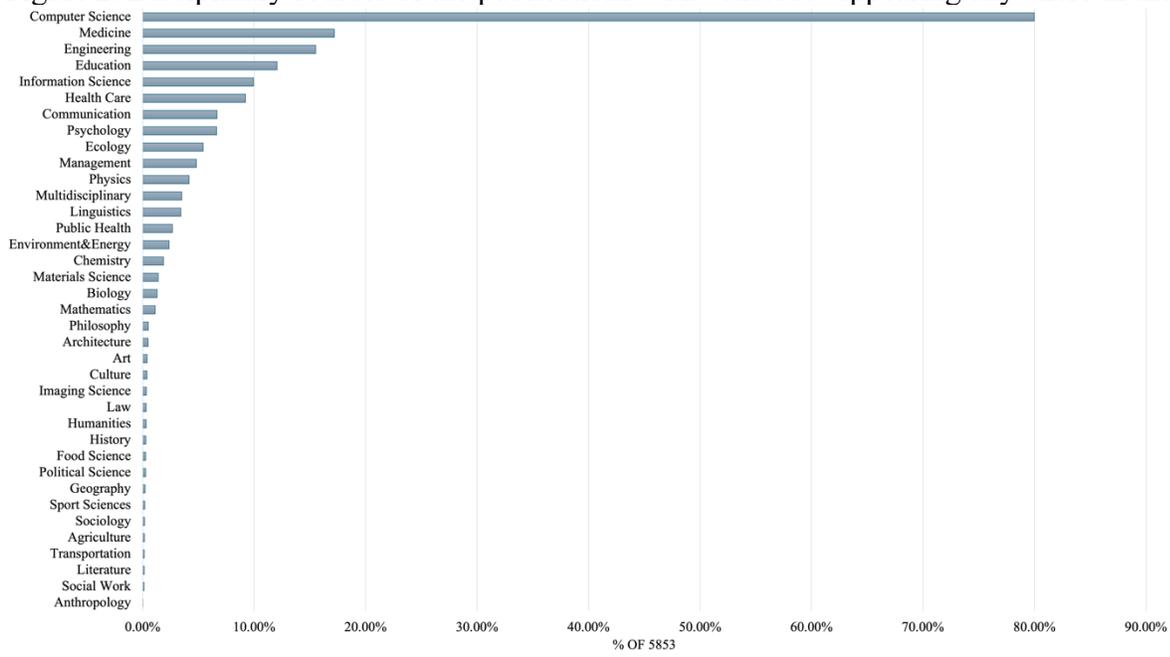

Source: Authors' own analysis of Web of Science data as of December 2024